\documentclass[useAMS,usenatbib]{mn2e}

\usepackage{graphicx,epsfig}

\def\gsim{\;\lower4pt\hbox{${\buildrel\displaystyle >\over\sim}$}\;}
\def\lsim{\;\lower4pt\hbox{${\buildrel\displaystyle <\over\sim}$}\;}
\def\grls{\;\lower4pt\hbox{${\buildrel\displaystyle >\over <}$}\;}

\title[Three Diagnostics of EECC Star-forming Cloud Model]
{Forming Protostars in Molecular Clouds with
\\ Shocked Envelope Expansion and Core Collapse}
\author[Yu-Qing Lou \& Yang Gao]
{Yu-Qing Lou $^{1,2,3}$\thanks{E-mail:louyq@tsinghua.edu.cn(Y-QL);
\hskip 2.5cm\hbox{}gaoyang-00@mails.tsinghua.edu.cn(YG)} and
Yang Gao$^{1}$\footnotemark[1]\\
$^1$Department of Physics and Tsinghua Centre for Astrophysics
 (THCA), Tsinghua University, Beijing 100084, China \\
$^2$National Astronomical Observatories of China, Chinese Academy
  of Sciences, A20, Datun Road, Beijing 100012, China \\
$^3$Department of Astronomy and Astrophysics,
  The University of Chicago, 5640 South Ellis Avenue, Chicago, IL 60637, USA }
\date{Accepted 2010 November 9. Received 2010 November 8; in original form 2009 December 26}

\begin{document}
\maketitle

\begin{abstract}
Spectral observations of molecular line profiles reveal the
  so-called `blue profiles' for double-peaked molecular
  lines with stronger blue and weaker red peaks as notable
  features for star-forming cloud core collapses under
  the self-gravity.
In contrast, $\sim 25-30$ per cent of observed molecular
  spectral line profiles in star-forming clouds or cores
  also show the so-called double-peaked `red profiles'
  with red peaks stronger than blue peaks.
  Gao \& Lou (2010) show that these unexplained
  `red profiles'
  can be signatures of global dynamics for envelope
  expansion with core collapse (EECC) within
  star-forming molecular clouds or cores.
We demonstrate here that spatially-resolved `red profiles' of
  HCO$^+$ (J$=1-0$) and CS (J$=2-1$) molecular transitions from the
  low-mass star-forming cloud core FeSt 1-457 together with its
  radial profile of column density inferred from dust extinction
  observations appear to reveal a self-similar hydrodynamic shock
  phase for global EECC.
Observed spectral profiles of C$^{18}$O (J$=1-0$)
  are also fitted by the same EECC model.
For further observational tests, the spatially-resolved profiles
  of molecular transitions HCO$^+$ (J$=3-2$) and CS (J$=3-2$)
  as well as the radial profiles of (sub)millimetre continuum
  emissions at three wavelengths of 1.2mm, 0.85mm and 0.45mm
  from FeSt 1-457 are also predicted.
\end{abstract}


\section{Introduction}

Gravitational collapses of molecular clouds and
  cloud cores lead to early formation of protostars therein
  \citep[e.g.][]{shu1977,shu1987,mclaughlin1997,mckee2002}.
For starless and star-forming molecular cloud cores in early
  phases,
  the central dense mass blob
  might not involve discs or outflows especially for those
  grossly spherical molecular clouds.
Recent theoretical development of self-similar
  hydrodynamic shock models involving the self-gravity
  reveals, among others, a novel plausible dynamic scenario
  of global envelope expansion with core collapse (EECC)
  \citep{lou2004,bianlou2005,lou2006,wang2008} with an
  expanding infall radius $R_{\rm inf}$ during early
  dynamic evolution stages of forming protostellar
  cores.
Together with other pertinent yet independent
  diagnostics and constraints,
  observations of various molecular emission spectral line
  profiles, especially those with high enough spatial and
  frequency resolutions and/or different line transitions
  of the same molecules, make it possible to identify the
  gross hydrodynamic and thermodynamic structures of
  star-forming molecular clouds with grossly spherical morphologies.
%

Major diagnostics to probe molecular cloud structures include
  spatially-resolved spectral profiles for various
  molecular line transitions
  \citep[e.g.][Fu, Gao \& Lou 2010]{zhou1993,tafalla2006},
  radial profiles of column densities inferred from dust
  extinctions of background stars behind a molecular cloud
  \citep[e.g. Alves et al. 2001;][]{
  kandori2005},
  radial profiles of (sub)millimetre continuum emissions
  \citep[e.g.][]{shirley2000,motte2001}
  etc.
Any of such separate observational diagnostics may be fitted with
  empirical and parametrized models on an individual basis to
  various extents as can be found in the literature.
In contrast and in order to remove potential model degeneracies,
  our strategy is to constrain the underlying cloud model
  using the
  hydrodynamics and by all available observational data simultaneously.
  We also make model
  predictions for specific observational tests.
We mainly focus on those molecular clouds which appear
  grossly spherical as our spherically symmetric hydrodynamics
  represents a first-order approximation in that direction.
For illustration, we choose the molecular cloud core FeSt 1-457
  for specific observational comparisons.
We also summarize the main results of our model comparisons
  with several observations for another star-forming cloud
  core L1517B as supporting evidence.

\begin{figure*}
\begin{center}
\epsfig{figure=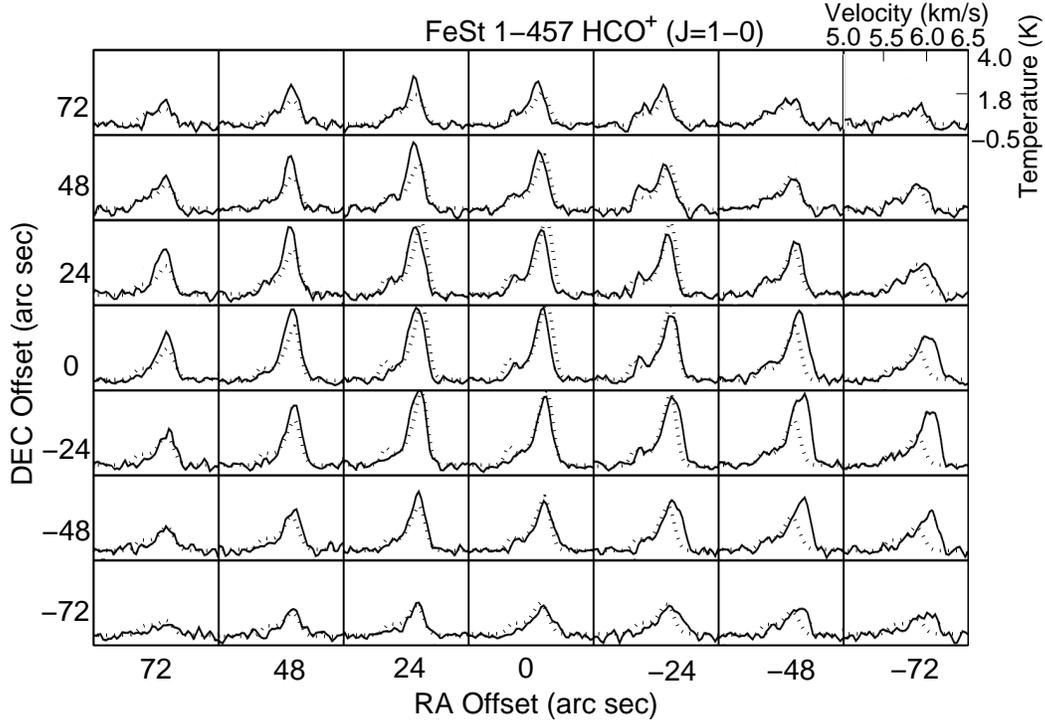,width=15.0cm,clip=}
\end{center}
\caption{The observed molecular line transition HCO$^+$ (J$=1-0$)
  in a spatially-resolved grid-map of star-forming molecular
  cloud core FeSt 1-457 (solid line profiles) superposed with our
  best-fit EECC shock dynamic model results (dotted line profiles).
The receding velocity of the entire cloud core FeSt 1-457
  along the line of sight is estimated to be $\sim 5.75$ km s$^{-1}$.
Each grid has the local receding velocity along the line
  of sight as the abscissa and antenna temperature as the
  ordinate (see the very upper right grid as a sample).
The observational data map of \citet{aguti2007} is centered at
  right ascension $\alpha_{J2000}=17^{\rm h} 35^{\rm m} 47.5^{\rm s}$
  and declination $\delta_{J2000}=-25^\circ 33' 2.0''$, with a grid
  spacing of $\sim 24''$ (corresponding to $\sim 3.0\times 10^3$~au
  at a distance of $\sim 130$~pc for the Pipe Nebula) and a maximum
  observed area of $168''\times 168''$ in the plane of the sky.
Molecular spectral line profile fittings are based on our general
  polytropic EECC shock model \citep{wang2008,gaolou2010}, by
  adopting a step function molecular abundance profile (see
  details in Table 1) and by performing the RATRAN computation.
Our model spectral line profiles are convolved with the beam width
  of $\sim 24''$ and multiplied by an IRAM 30m telescope mean
  beam efficiency of $\sim 0.75$. \label{fig:hco}}
\end{figure*}

Regarding the molecular line diagnostics, there
  are two major asymmetric double-peak molecular spectral
  line profiles oft-observed in star-forming clouds,
  viz., the so-called `blue profiles' with stronger blue-shifted
  peaks than red-shifted peaks and the so-called `red profiles'
  with stronger red-shifted peaks than blue-shifted peaks.
  The level of such spectral line profile asymmetries can change
  in a gradual manner for spatially-resolved observations.
While the observed `blue profiles' of molecular spectral lines are
  modelled and generally interpreted as characteristic signatures
  of radial infalls in molecular clouds \citep{zhou1993,gao2009},
  the somewhat ignored yet wide presence of unexplained `red
  profiles' \citep[e.g.][]{mardones1997,evans2003,velusamy2008}
  remains a conceptual gap in our understanding of star-forming
  molecular cloud dynamics.
Moreover, sometimes `blue profiles' and `red profiles' coexist
  in molecular cloud cores for the same line transition
  \citep[e.g.][]{tafalla2000}, and/or for different line
  transitions \citep[e.g.][]{tafalla2006}.
For the possible interpretation of `red profiles' of
  molecular spectral lines alone, one may think of several
  mechanisms, e.g. rotations \citep[e.g.][]{difrancesco2001},
  bipolar outflows inclined close to the plane of the sky
  \citep[e.g.][]{fiege1996}, turbulence
  \citep[e.g.][]{ossenkopf2002} and thermal
  pulsations \citep[e.g.][]{lada2003}.

For FeSt 1-457 (see Fig. 1), disc rotation is unlikely, as this
  cloud core is most likely evolving in an early phase.
A global cloud core rotation of FeSt 1-457 may be excluded by
  spatially-resolved spectral line observations; i.e., the
  currently data do not show such evidence.
At present, these mechanisms just mentioned have not been
  shown to also account for cloud density profiles and radial
  profiles of (sub)millimetre continuum emissions simultaneously
  within a consistent model framework.
Alternatively, it was suggested that `red profiles' in certain
  star-forming sources may be caused by clouds involving EECC
  \citep{lou2004,thompson2004}.
It is thus crucial to reproduce the observed `red profiles'
  using our general polytropic EECC shock models via
  radiative transfer calculations \citep{gaolou2010}.
Meanwhile, we require the model density profile to be consistent
  with the observational inference of dust extinctions and further
  predict other spectral line profiles as well as radial profiles of
  (sub)millimetre continuum emissions as future observational tests.

Four main contributions of our model analysis in this paper are
  as follows.
First, for the grossly spherical molecular cloud
  core FeSt 1-457, we construct a self-similar
  general polytropic hydrodynamic EECC shock model
  \citep{lou2004,lou2006,wang2008},
  and radiative transfer calculations are performed
  to reproduce the observed `red profiles' and other spectral
  profiles of molecular transition lines in the cloud core.
Secondly, we advance the conceptual framework of constraining
  the molecular cloud dynamics using several available diagnostics
  simultaneously in order to solve this challenging issue in
  molecular astronomy.
For FeSt 1-457, the column density profile of our EECC shock
  dynamics is also sensibly fitted with the observationally
  inferred column density profile by the dust extinction data
  \citep[e.g.][]{kandori2005}.
Thirdly, while no data for (sub)millimetre continua are available
  for FeSt 1-457 at present,
  for scenario consistency and further observational tests,
  we predict radio continuum emissions at three wavelengths
  of 1.2 mm, 0.85 mm and 0.45 mm.
For the same reason, the spatially-resolved spectral profiles of
  molecular line transitions HCO$^+$ (J$=3-2$) and CS (J$=3-2$)
  are also predicted for observational tests.
Finally, we suggest \citep{gaolou2010} that such EECC shock phase
  may evolve from damped nonlinear radial pulsations of molecular
  cloud core within one or two cycles.
Corresponding to different phases of such damped large-scale
  radial pulsations, different dynamic phases may give rise
  to distinct diagnostics.

\section{Star-Forming Molecular\\ \ \ \ \
Cloud Core FeSt 1-457}

As a chosen candidate, the molecular cloud core FeSt 1-457 appears
  to be a grossly spherical dark globule in the Pipe Nebula at
  a distance of $\sim 130$~pc away (e.g. Lombardi et al. 2006).
Without detecting a star, this starless cloud core in the early
  phase of protostar formation has a diameter of $\sim 35,000$ au
  with an angular diameter $\sim 4.5'$ derived from the optical
  images \citep[e.g.][]{aguti2007}.
The spectral data from \citet[][]{aguti2007} for
  spatially-resolved spectral line profiles of three molecular transitions
  HCO$^+$ (J$=1-0$), CS (J$=2-1$) and C$^{18}$O (J$=1-0$) (Figs.
  \ref{fig:hco} and \ref{fig:csfit}) were acquired in 2003 by the
  30-metre millimetre-wave telescope of Institut de Radioastronomie
  Millim{\'e}trique (IRAM) on Pico Veleta near Granada in Spain.
The spatial resolution of this spectral data set is $\sim 24''$
  for the beam width in the frequency band $90-110$~GHz with a
  mean beam efficiency of $\sim 0.75$ at uncertainties $\lsim 4$
  per cent (e.g. Greve et al. 1998).
The dust extinction data of FeSt 1-457 for the inferred radial
  profile of column density with a $\sim 30''$ beam resolution
  are also available together with error bars
  \citep[e.g.][and see the upper panel of our Fig.
  \ref{fig:continuum}]{kandori2005}.

Global expansions in the outer layer or envelope of this globule
  are hinted by asymmetric molecular spectral line profiles and cloud
  oscillations were proposed to be the origin of such expansions
  \citep[e.g.][]{aguti2007}.
In our advocated scenario, such large-scale nonlinear cloud radial
  pulsations cannot persist forever but may initiate core collapses
  and envelope expansions within one or two cycles and hence may lead
  to a self-similar EECC shock phase \citep{lou2006,gaolou2010}.
We demonstrate that the global EECC shock phase seem to persist in
  FeSt 1-457 by molecular spectral line profile fits using the
  spherically symmetric general polytropic self-similar dynamic
  model \citep{wang2008}.
Meanwhile, we also use the dust extinction data
  \citep{kandori2005} of FeSt 1-457 for the radial profile of column
  density with a $\sim 30''$ beam resolution to constrain our EECC
  shock hydrodynamics (Fig. \ref{fig:continuum}).
Finally for future observational tests, we predict the
  spatially-resolved spectral profiles of molecular line transitions
  HCO$^+$ (J$=3-2$) and CS (J$=3-2$) as well as three radial profiles
  of millimetre continua at wavelengths 1.2mm, 0.85mm and 0.45mm with
  estimated error ranges using the density and temperature profiles
  of the same underlying EECC shock dynamics (Fig. \ref{fig:hco32} and \ref{fig:continuum}),
  as have been already detected in other protostar-forming molecular
  cloud cores \citep[e.g.][]{shirley2002}.

\section{EECC Shock Hydrodynamics and Formation
   of Red Profiles for Molecular Spectral Lines}

\subsection{Self-similar Hydrodynamic Models
\\ \qquad\ and EECC Shock Solutions}

EECC self-similar hydrodynamic shock solutions have been
  investigated with different equations of state (EoS)
  in recent years \citep{lou2004,lou2006,wang2008}.
We here adapt the magnetohydrodynamic (MHD) model formulation
  with general polytropic EoS in \citet{wang2008} but without
  the magnetic field (i.e. setting the magnetic field parameter
  $h=0$ in that paper).

Our model formulation begins with the ideal hydrodynamic
  partial differential equations (PDEs) for a spherically
  symmetric molecular cloud in spherical polar coordinates
  $(r,\ \theta,\ \phi)$, namely
  \begin{equation}
  {{\partial\rho}\over{\partial t}}
  +{1\over{r^2}}{{\partial}\over{\partial r}}(r^2\rho u)=0\ ,
  \label{Equ:mass1}
  \end{equation}
  \begin{equation}
  {{\partial u}\over{\partial t}}+u{{\partial u} \over {\partial
  r}}=-{1\over{\rho}}{{\partial p}\over {\partial
  r}}-{{GM}\over{r^2}}\ , \label{Equ:force}
  \end{equation}
  \begin{equation}
  {{\partial M}\over{\partial t}}+u{{\partial M}\over{\partial
  r}}=0\ ,\qquad\qquad\qquad {{\partial M}\over{\partial r}}=4\pi
  r^{2}\rho\ ,\label{Equ:mass2}
  \end{equation}
  \begin{equation}
  \left(\frac{\partial}{\partial t}+u\frac{\partial}{\partial r}\right)
  \left({\rm ln}\frac{p}{\rho^{\gamma}}\right)=0\ ,\label{Equ:entropy}
  \end{equation}
  where, as functions of both radius $r$ and time $t$, dependent
  variables $u$, $\rho$, $M$, and $p$ are the radial flow velocity,
  mass density, enclosed mass within radius $r$ at time $t$ and
  thermal gas pressure, respectively; $G$ is the gravitational
  constant and $\gamma$ is the polytropic index.
These nonlinear hydrodynamic PDEs describe the conservations of
  mass (eq (\ref{Equ:mass1})), radial momentum (eq
  (\ref{Equ:force})) and specific entropy along
  streamlines (eq (\ref{Equ:entropy})).

Self-similar solutions to nonlinear PDEs
  (\ref{Equ:mass1})$-$(\ref{Equ:entropy}) by the following
  self-similar transformation can be constructed, namely
\begin{eqnarray}
  r=k^{1/2}t^n x\ ,
  \label{equ:radius}
\end{eqnarray}
\begin{eqnarray}
  u=k^{1/2}t^{n-1}v(x)\ ,\qquad\qquad
  \rho=\frac{\alpha(x)}{4\pi Gt^2}\ ,
  \label{equ:varu}
\end{eqnarray}
\begin{eqnarray}
  M=\frac{k^{3/2}t^{3n-2}m(x)}{(3n-2)G}\ ,\qquad
  p=\frac{kt^{2n-4}\alpha(x)^{\gamma}m(x)^q}{4\pi G}\ ,\
  \label{equ:varp}
\end{eqnarray}
where $x$ is the self-similar independent variable combining
  $r$ and $t$ in a special manner, and $v(x)$, $\alpha(x)$ and $m(x)$
  are functions of $x$ only and represent the dimensionless reduced
  radial flow velocity, mass density and enclosed mass, respectively;
  and $n$ and $\gamma$ in these self-similar transformations are two
  additional scaling indices allowed, with the algebraic relation
  $q=2(n+\gamma-2)/(3n-2)$.
Here $k$ is a parameter related to the sound speed
  within the molecular cloud.
Physically, a positive enclosed mass $M$ requires the inequality
  $n>2/3$ in the similarity transformation above.
For $n+\gamma=2$ and thus $q=0$ as in this paper unless otherwise
  stated, we reduce to the conventional polytropic hydrodynamics
  \citep{wang2007,wang2008} as an important subcase.
Other important physical variables, namely the gas
  thermal temperature (invoking the ideal gas law) and the
  central mass accretion rate in the molecular cloud centre are
  \begin{equation}
  T\equiv\frac{p}{k_{\rm B}\rho/(\mu m_{\rm H})}=
  \frac{\mu m_{\rm H}}{k_B}kt^{2(n-1)}\alpha(x)^{\gamma-1}m(x)^q\ ,
  \label{equ:temp}
  \end{equation}
  \begin{equation}
  \dot{M_0}=k^{3/2}t^{3(n-1)}m_0/G\ ,\label{equ:massaccretion}
  \end{equation}
  respectively.
Here $m_0\equiv m(0)$ is the reduced central enclosed mass,
  $k_{\rm B}$ is Boltzmann's constant, $m_{\rm H}$ is the atomic
  hydrogen mass, and $\mu$ is the mean molecular weight setting
  to be 2 in our model analysis for star-forming gas cloud
  consisting mainly of molecular hydrogen H$_2$.

By adopting self-similar transformation equations
  (\ref{equ:radius})$-$(\ref{equ:varp}), nonlinear hydrodynamic
  PDEs (\ref{Equ:mass1})$-$(\ref{Equ:entropy}) can be readily
  cast into two coupled nonlinear ordinary differential
  equations (ODEs) which can be solved by a combination of
  analytical and numerical techniques \citep[see, e.g.][]{wang2008}.
These two coupled nonlinear ODEs are
\begin{eqnarray}
\bigg[\bigg(2-n+\frac{3n-2}{2}q\bigg)
\alpha^{1-n+\frac{3nq}{2}}x^{2q}(nx-v)^q-(nx-v)^2\bigg]
\nonumber \\
\times\frac{d\alpha}{dx}=2\frac{(x-v)}{x} \alpha\bigg[(nx-v)
+q\alpha^{1-n+\frac{3nq}{2}}x^{2q}(nx-v)^{q-1}\bigg]
\nonumber \\
-\alpha\bigg[(n-1)v+\frac{(nx-v)}{(3n-2)}\alpha
\qquad\qquad\qquad\qquad\qquad\qquad \nonumber
\\
+q\alpha^{1-n+\frac{3nq}{2}}x^{2q-1}(nx-v)^{q-1}(3nx-2v)\bigg]\ ,
\qquad \label{equ:ode1}
\end{eqnarray}
\begin{eqnarray}
\bigg[\bigg(2-n+\frac{3n-2}{2}q\bigg)
\alpha^{1-n+\frac{3nq}{2}}x^{2q}(nx-v)^q-(nx-v)^2\bigg]
\nonumber \\
\times\frac{dv}{dx}=2\frac{(x-v)}{x}\bigg(2-n+\frac{3nq}{2}\bigg)
\alpha^{1-n+\frac{3nq}{2}}x^{2q}(nx-v)^q
\nonumber \\
-(nx-v)\bigg[(n-1)v+\frac{(nx-v)}{(3n-2)}\alpha \qquad\qquad\qquad
\nonumber
\\
+q\alpha^{1-n+\frac{3nq}{2}}x^{2q-1}(nx-v)^{q-1}(3nx-2v)\bigg]\ .
\label{equ:ode2}
\end{eqnarray}
In the above mathematical derivations, we also make use of the
   following algebraic relation
\begin{equation}\label{Louadd}
m=\alpha x^{2}(nx-v)\
\end{equation}
from the mass conservation (\ref{Equ:mass1}) and
  (\ref{Equ:mass2}).
Once $\alpha(x)$ and $v(x)$ are known, we then readily
  obtain the reduced enclosed mass $m(x)$.
Two sets of analytical asymptotic similarity solutions
  for $x\rightarrow +\infty$ and $x\rightarrow 0^{+}$
  respectively serve as the boundary conditions in
  solving two coupled nonlinear ODEs (\ref{equ:ode1})
  and (\ref{equ:ode2}).
Among other possible solutions, we have for the
  $x\rightarrow +\infty$ regime
  \begin{eqnarray}
  \alpha=Ax^{-{2}/{n}},
  \qquad\qquad\qquad\qquad\qquad
  \qquad\qquad\qquad\qquad\nonumber \\
  v=\bigg[-\frac{nA}{(3n-2)}+2(2-n)n^{q-1}A^{1-n+3nq/2}\bigg]
  x^{{(n-2)}/{n}}\nonumber\\
  +Bx^{{(n-1)}/{n}}\ ,\label{Equ:infinity}
  \end{eqnarray}
  and for the $x\rightarrow 0^{+}$ regime,
  the central free-fall solution
  \begin{equation}
  v=-\bigg[\frac{2m_{0}}{(3n-2)x}\bigg]^{1/2}\ ,\qquad\ \
  \alpha=\bigg[\frac{(3n-2)m_{0}}{2x^{3}}\bigg]^{1/2}\ .
  \label{Equ:zero1}
  \end{equation}
Here, the mass parameter $A$ and the velocity parameter
  $B$, describing the asymptotic boundary condition as
  $x\rightarrow +\infty$, are two constants of integration
  and are important parameters to characterize the global
  self-similar dynamic solution.
Meanwhile, the reduced enclosed mass $m_0\equiv m(0)$
  describes the central point mass embedded deep in the cloud core.
Then our global self-similar dynamic solutions can be constructed
  by straightforward numerical integrations using the standard
  fourth-order Runge-Kutta scheme and by specifying boundary
  conditions (\ref{Equ:infinity}) and (\ref{Equ:zero1}).
In the absence of a shock, $A$ and $B$ are specified and $m_0$ is
  determined accordingly; i.e. only two are independent among the
  three parameters.
For solutions with shocks, the shock location is another parameter
  and discontinuity across a shock front is introduced; global
  solutions can be still constructed by simply requiring the conservation
  laws of mass, radial momentum and energy across the upstream and
  downstream sides of the shock front (see \citet{wang2008} for more
  details regarding the hydrodynamic formulation).

\subsection{The EECC hydrodynamic Shock Solution
\\ \qquad
  Adopted for the Cloud Core FeSt 1-457}

%
\begin{figure}
\begin{center}
\epsfig{figure=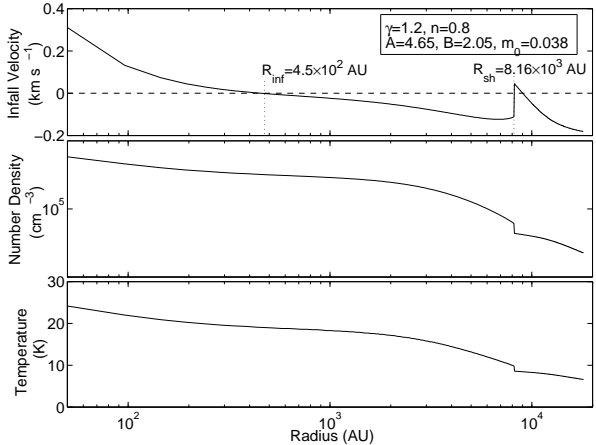,width=9cm,clip=}
\end{center}
\caption{Radial profiles of infall speed $-u(r,\ t)$, H$_2$ number
density $N(r,\ t)$ and thermal temperature $T(r,\ t)$ for the
polytropic EECC shock dynamic phase of a model cloud for FeSt 1-457
with an evolutionary timescale $t\sim 2.8\times 10^5$ yr. This EECC
shock dynamic model has an expanding infall radius $R_{\rm
inf}=4.5\times 10^2$~au (top panel), a shock radius $R_{\rm
sh}=8.16\times 10^3$~au (all three panels) and an outer cloud
boundary of $R_{\rm out}=1.80\times 10^4$~au. The infall radius
$R_{\rm inf}$ and shock radius $R_{\rm sh}$ expand at speeds of
$\sim 0.015~{\rm km~s^{-1}}$ and $\sim 0.27~{\rm km~s^{-1}}$,
respectively. The top panel shows that the cloud core collapses
inside $R_{\rm inf}$ while the envelope outside $R_{\rm inf}$
expands outwards with typical gas flow speeds $\lsim 0.1~{\rm
km~s^{-1}}$. The middle and bottom panels show a number density
increase by a factor $\sim 10^3$ and a temperature increase by a
factor $\gsim 3$ times towards the core centre where a point mass
blob of $M_0=0.0202M_{\odot}$ resides. \label{fig:dynamics} }
\end{figure}

In our model analysis, various EECC shock dynamic solutions
  are extensively explored in order to plausibly represent
  the dynamic state of FeSt 1-457 by adopting model fitting
  procedures detailed in Section 4.
Finally after an extensive search, a sample EECC shock dynamic
  solution with polytropic index $\gamma=1.2$, scaling parameter $n=0.8$
  and three coefficients of asymptotic conditions $A=4.65$, $B=2.05$
  and $m_0=0.038$ is chosen\footnote{The outgoing shock radius
  and thus shock travelling speed at current epoch $t$ are
  then determined.}
  to fit the molecular spectral line profile data and the column
  density radial profile inferred from dust extinction observations.
Two relevant physical scalings
  \citep[see subsection 2.3 of][]{gaolou2010} of the length scale
  and number density scale are specifically chosen below as
\begin{equation}
  k^{1/2}t^n=4.8\times 10^3~{\rm au}\ ,
  \label{equ:scale1}
\end{equation}
\begin{equation}
  (4\pi G\mu m_{\rm H}t^2)^{-1}
  =2.7\times 10^4~{\rm cm^{-3}}\ ,
  \label{equ:scale2}
\end{equation}
  with the mean molecular weight $\mu= 2$.

Our EECC hydrodynamic shock solution offers the global radial
  structural profiles of radial infall velocity $u(r,\ t)$,
  number density $N(r,\ t)$ and thermal gas temperature $T(r,\ t)$
  (see Fig. \ref{fig:dynamics}); the cloud temperature $T(r,\ t)$
  increases towards the centre and an infall radius
  $R_{\rm inf}=4.5\times 10^2~{\rm au}$ expanding at a speed of
  $\sim 0.015~{\rm km~s^{-1}}$ separates the inner core collapse
  domain ($u<0$) and the outer envelope expansion domain ($u>0$).
An outgoing spherical shock is at radius $R_{\rm sh}=8.16\times
  10^3$~au with a shock travelling speed of $\sim 0.27~{\rm km~s^{-1}}$
  and the outer cloud core radius is at $R_{\rm out}=1.80\times 10^4$~au.
Following the chosen scales by expressions (\ref{equ:scale1}) and
  (\ref{equ:scale2}), the sound parameter $k$ values across a
  shock front are different; the upstream (pre-shock) side has
  $k^{1/2}=k_{\rm u}^{1/2}=48.7~{\rm km~s^{-0.8}}$ and the downstream
  (post-shock) side has $k^{1/2}=k_{\rm d}^{1/2}=49.1~{\rm km~s^{-0.8}}$.
Correspondingly (also as a result of different densities in the
  upstream and downstream sides, see eq. (8) in Lou \& Gao 2006),
  the sound speed $c$ across the outgoing shock front is
  different:
  $c_{\rm u}=0.133~{\rm km\ s}^{-1}$ and
  $c_{\rm d}=0.143~{\rm km\ s}^{-1}$.
This EECC shock dynamic model describes a molecular cloud
  core in the early phase of star formation (a timescale of
  $t\sim 2.8\times 10^5~{\rm yr}$) with a central mass accretion
  rate of $\dot{M}_0\sim 0.69\times 10^{-7}~M_\odot~{\rm yr}^{-1}$
  and a central point mass blob of $M_0=0.0202~M_\odot$.

For the identification of a sensible self-similar EECC shock model
  for FeSt 1-457, we have explored different independent parameters
  within the model framework, namely the polytropic index $\gamma$,
  the mass and velocity parameters $A$ and $B$ for the asymptotic
  boundary condition at large $x$, and a self-similar radius $x_1$
  describing the shock location and thus the shock speed.
Then the other two indices, $n$ and $\gamma$ are naturally fixed
  under the condition of conventional polytropic case with
  $n+\gamma=2$ and $q=0$;
and the inner reduced point mass $m_0$ is also
  determined by the global trend of the specific EECC shock
  solution as $A$, $B$ and shock location $x_1$ are chosen.
There are additional independent parameters in the scaling
  process of the self-similar model, namely the length scale
  (\ref{equ:scale1}) and the number density scale (\ref{equ:scale2}).
The current set of independent parameters for the EECC shock
  dynamic model, without degeneracy as we know, sensibly describes
  physical properties within molecular cloud core FeSt 1-457
  when we compare the hydrodynamic shock cloud model with currently
  available observations as described in Section 4 presently.
We further venture to make pertinent predictions based on our
  theoretical model analysis for future observational tests.

\subsection{Formation of Red Asymmetric Profiles for
  Molecular Spectral Line Transitions}

%
\begin{figure}
\begin{center}
\epsfig{figure=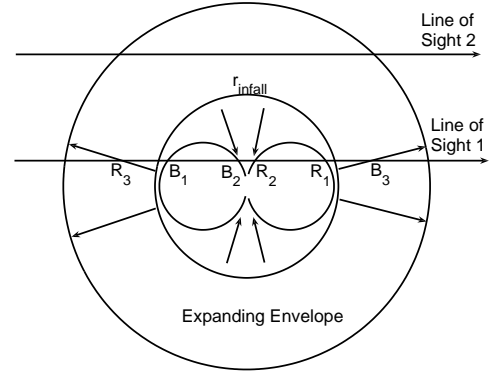,width=9cm,clip=}
\end{center}
\caption{A schematic illustration of the envelope expansion with
  core collapse (EECC) scenario for generating red profiles of
  asymmetric molecular line transitions.
The cloud consists of an outer expanding envelope and an inner
  collapsing core, separated by an outgoing spherical surface
  at $r=r_{\rm infall}$ in a self-similar radial expansion.
The loci (i.e. the two connected partial ovals inside the
  infall radius $r_{\rm infall}$) and two straight arrow lines
  pointing right of particularly chosen constant projected
  line-of-sight velocity components are shown as examples.
Line of Sight 1 passes through the core of central infall as well
  as the outer expanding envelope, that intercepts the ovals and
  straight arrow lines at three points B$_1$, B$_2$ and B$_3$ for
  the same approaching projected velocity components and at three
  points R$_1$, R$_2$ and R$_3$ for the same receding projected
  velocity components.
Sufficiently far away from the cloud centre, Line of Sight 2
  passes completely through the outer expanding envelope.
  \label{fig:schematic}
  }
\end{figure}

Why does our EECC dynamic model lead to the red profiles of
  molecular spectral lines?
Before formal radiative transfer calculations, we would present
  a schematic analysis in Fig. \ref{fig:schematic}.
Line of Sight 1, which is close to the cloud core centre,
  intercepts the ovals and straight arrow lines at points
  labelled `B' and `R' for the same approaching and receding
  projected velocity components, respectively.
The intensity of an optically thick transition at a certain
  frequency will be dominated by emission from the nearest
  component to the observer, i.e. R$_1$ and B$_3$ along Line
  of Sight 1, because of the absorption and scattering along
  the path.
Assuming the local thermal equilibrium (LTE) for simplification,
  when the gas temperature increases towards the core centre (as
  for the case in the bottom panel of Fig. \ref{fig:dynamics}),
  emission from point R$_1$ will be stronger than that from
  point B$_3$, which then gives rise to the asymmetric red
  profile feature in double-peaked molecular line profiles.
On the other hand, as Line of Sight 2 passes completely through
  the expanding envelope, no `red profile' feature would be
  detected in this situation.

For a more detailed analysis, we would first
  invoke the following radiative transfer equation
  \citep[e.g.][]{chandrasekhar1960,rybicki1978}, namely
  \begin{equation}
  \frac{d}{d\tau}I_\nu =-I_\nu+S_\nu\ ,\label{equ:RTequ}
  \end{equation}
  with $\tau$ being the optical depth, $S$ being the source
  function and $I$ being the radiative intensity; subscript $\nu$
  denotes that these variables are all at a particular frequency $\nu$.
Then the radiative intensity at blue and red shifted frequencies are
  \begin{eqnarray}
  I_{\rm
  blue}=S_1(1-e^{-\tau_1})e^{-(\tau_2+\tau_3)}
  \qquad\qquad\qquad\qquad\qquad\\ \nonumber
  \qquad\qquad\qquad +S_2(1-e^{-\tau_2})e^{-\tau_3}
  +S_3(1-e^{-\tau_3})\ ,
  \
  \label{equ:blue} \\
  I_{\rm
  red}=S_3(1-e^{-\tau_3})e^{-(\tau_1+\tau_2)}
  \qquad\qquad\qquad\qquad\qquad\\ \nonumber
  \qquad\qquad\qquad +S_2(1-e^{-\tau_2})e^{-\tau_1}
  +S_1(1-e^{-\tau_1})\ ,
  \label{equ:red}
  \end{eqnarray}
respectively, where subscripts 1, 2 and 3 denote
  variables at three respective positions along Line
  of Sight 1 shown in Fig. \ref{fig:schematic}.
Therefore the difference between the blue-shifted
  and red-shifted intensities $I_{\rm blue}$ and
  $I_{\rm red}$ around this frequency is
  \begin{eqnarray}
  I_{\rm blue}-I_{\rm red}=(S_3-S_1)
  [1+e^{-(\tau_1+\tau_2+\tau_3)}]
  \qquad\qquad \\ \nonumber
  \qquad
  +(S_2-S_3)[e^{-\tau_3}-e^{-(\tau_1+\tau_2)}]
  \qquad \\ \nonumber
  \qquad\qquad\qquad\qquad\qquad
  +(S_1-S_2)[e^{-\tau_1}-e^{-(\tau_2+\tau_3)}]\ .
  \label{equ:diffintensity}
  \end{eqnarray}
Under the LTE condition and the Rayleigh-Jeans approximation
  for a blackbody source function \citep[e.g.][]{rybicki1978},
  intensities can be expressed in units of brightness temperatures.
Therefore the difference between the blue-shifted and red-shifted
  peak brightness temperatures $T_{\rm blue}$ and $T_{\rm red}$ is
  \begin{eqnarray}
  T_{\rm blue}-T_{\rm red}=
  (T_3-T_1)[1+e^{-(\tau_1+\tau_2+\tau_3)}]
  \qquad\qquad \\ \nonumber
  \qquad +(T_2-T_3)[e^{-\tau_3}-e^{-(\tau_1+\tau_2)}]
  \qquad \\ \nonumber
  \qquad\qquad\qquad\qquad\qquad
  +(T_1-T_2)[e^{-\tau_1}-e^{-(\tau_2+\tau_3)}]\
  .\label{equ:difftemperature}
  \end{eqnarray}
We can immediately conclude from equation (21)
  that in an optically thick cloud environment (i.e. $\tau_1\gg 1$,
  $\tau_2\gg1$, $\tau_3\gg 1$), the difference in peak brightness
  temperatures $T_{\rm blue}-T_{\rm red}\approx T_3-T_1$.
Since the gas temperature generally tends to increase towards the
  cloud centre, i.e. $T_2>T_1>T_3$, we then have
  $T_{\rm blue}-T_{\rm red}<0$ corresponding to the red-shifted
  peak intensity being greater than the blue-shifted peak
  intensity (i.e. an asymmetric red profile).
In the opposite limit of optically thin regime (i.e. $\tau_1\ll
  1$, $\tau_2\ll 1$, $\tau_3\ll 1$), the difference in brightness
  temperatures $T_{\rm blue}- T_{\rm red}\approx 0$, indicating
  that we will get equal peak intensities in blue- and red-shifted
  frequencies.
When the blue- and red-shifted frequencies change (i.e. the loci
  of equal-projection velocity changes), all foregoing analytical
  results remain valid.
Therefore we can apply these results to the entire blue- and
  red-shifted peaks in molecular spectral line profiles,
  and the following conclusion can be reached:
double peaked molecular line profiles with stronger red peaks
  than blue peaks (i.e. red profiles) form under the optical
  thick condition in a dynamic EECC cloud core.
If the outer expanding envelope is replaced by a static envelope,
  we would then have blue profiles for molecular transitions for
  optically thick lines along the Line of Sight 1 in Fig.
  \ref{fig:schematic} (see also Gao, Lou \& Wu 2009 for more details).

\section{Data Fitting of EECC Shock Model }

\subsection{Molecular Spectral Line Profiles}

Adopting this EECC shock dynamic solution to model FeSt 1-457, we
  perform radiative transfer calculations for HCO$^+$ (J$=1-0$)
  line profiles using the RATRAN code \citep{hogerheijde2000}
  with spherical symmetry.
The Doppler b parameter in the code is taken as 0.1 km
  s$^{-1}$, corresponding to an intrinsic line broadening of
  0.17 km s$^{-1}$.
We introduced twelve shells with different thickness
  (more concentrated towards the cloud central region to
  account for the rapid variation of physical properties
  there) in performing the calculations.
The dust emissivity follows the model of \citet{ossenkopf1994}
  with bare ice mantles and with $\sim 10^5$ years of growth.
The abundance ratio of HCO$^+$ to H$_2$ in this cloud
  core is listed in Table 1.
Abundance variations in radius $r$ would affect molecular
  line profiles \citep[e.g.][]{tsamis2008}.
Here a constant abundance with a central fractional drop
  (i.e. a step function) appears to be a sensible
  prescription \citep[see also][]{evans2005,tafalla2006}.
Physically, molecular abundance ratios vary with the initial
  metal compositions and the evolution history of a molecular cloud
  \citep[e.g.][]{tsamis2008}; and the central abundance drop is
  caused by adhesions of molecules onto dust grains in a relatively
  higher temperature environment \cite[e.g.][]{tafalla2006}.
In the molecular spectral line profile fittings here, abundance
  values and the radii of abundance holes (see Table 1) in the
  step function serve as additional fitting parameters to
  achieve reasonable agreements between our model results
  and the observational data.

Given the beam width of $\sim 24''$, we average our model spectral
  line profiles over a beam area of $\sim 24''\times 24''$ and
  multiply a beam efficiency of $\sim 0.75$ for the IRAM 30m
  telescope (dotted curves in Fig. 1) to fit the line profile data.
Fairly reasonable overall fits of model line profiles to those
  actually observed `red profiles'
  \citep[solid curves in Fig. \ref{fig:hco} from][]{aguti2007}
  support EECC shock dynamic structures (Fig. \ref{fig:dynamics})
  for the dark globule FeSt 1-457.
Red profiles and red-skewed single peak line profiles
  in Fig. \ref{fig:hco} are encouraging indicators of expansive
  motions present within this cloud core.
In particular, the central $3\times 3$ grids show good agreements
  between observations and the results of spherical EECC shock
  model, suggesting that the spherical global expansion is at
  least a sensible first-order dynamic pattern characterizing
  the molecular cloud core FeSt 1-457.
Certain deviations of spectral line profiles towards the right
  bottom grids in Fig. \ref{fig:hco} then suggest that there
  might be a beamed outflow or certain rotations [as have been
  discussed in \citet{gaolou2010}] overlapped with the global
  envelope expansive motions described by the spherically
  symmetric model.
We may infer from the differences between model results and
  observations in Fig. \ref{fig:hco} that this additional
  asymmetric motions are of the order of
  $\sim 0.1{\ \rm km~s^{-1}}$.

\begin{table}
 \centering
 \begin{minipage}{80mm}
 \caption{Molecular line transition properties: $\nu_0$ is the
 rest-frame central frequency of the line transition;
 X$_0$ is the molecular abundance with respect to H$_2$ molecules
 in the cloud core FeSt 1-457; and $r_{\rm h}$ is the radius of
 the central abundance hole, within which a fractional abundance
 drop of $10^{-4}$ is presumed.$^a$}
 \begin{center}
  \begin{tabular}{@{}llllll@{}}
  \hline
 ${\rm Molec.}$&${\rm Trans.}$&$\nu_0{\rm(GHz)}$&X$_0$&
 $r_{\rm h}{\rm (cm)}$&{\rm Note}\\
 \hline
HCO$^{+}$&J$=1-0$&89.19 &2$\times 10^{-9}$&7.5 $\times 10^{16}$&    \\
HCO$^{+}$&J$=3-2$&267.56&2$\times 10^{-9}$&7.5 $\times 10^{16}$&$^b$\\
CS       &J$=2-1$&97.98 &3$\times 10^{-9}$&1.05$\times 10^{17}$&    \\
CS       &J$=3-2$&146.97&3$\times 10^{-9}$&1.05$\times 10^{17}$&$^b$\\
C$^{18}$O&J$=1-0$&109.78&1$\times 10^{-7}$&6.0 $\times 10^{16}$&$^c$\\
\hline
\end{tabular}
\end{center}
$^a$ Frequency values are taken from the LAMDA
  \citep[e.g.][]{schoier2005} and abundance values
  are close to those of \citet{tafalla2006} with
  slight variations.
\newline
$^b$ Theoretical model predictions
  for future observational tests.
\newline
$^c$ The abundance of C$^{18}$O is enhanced to ten times the
  value of $X_0$ for cloud radius $r>1.80\times 10^{17}$~cm.
  \label{Table:dynamics}
\end{minipage}
\end{table}

%
\begin{figure}
\begin{center}
\epsfig{figure=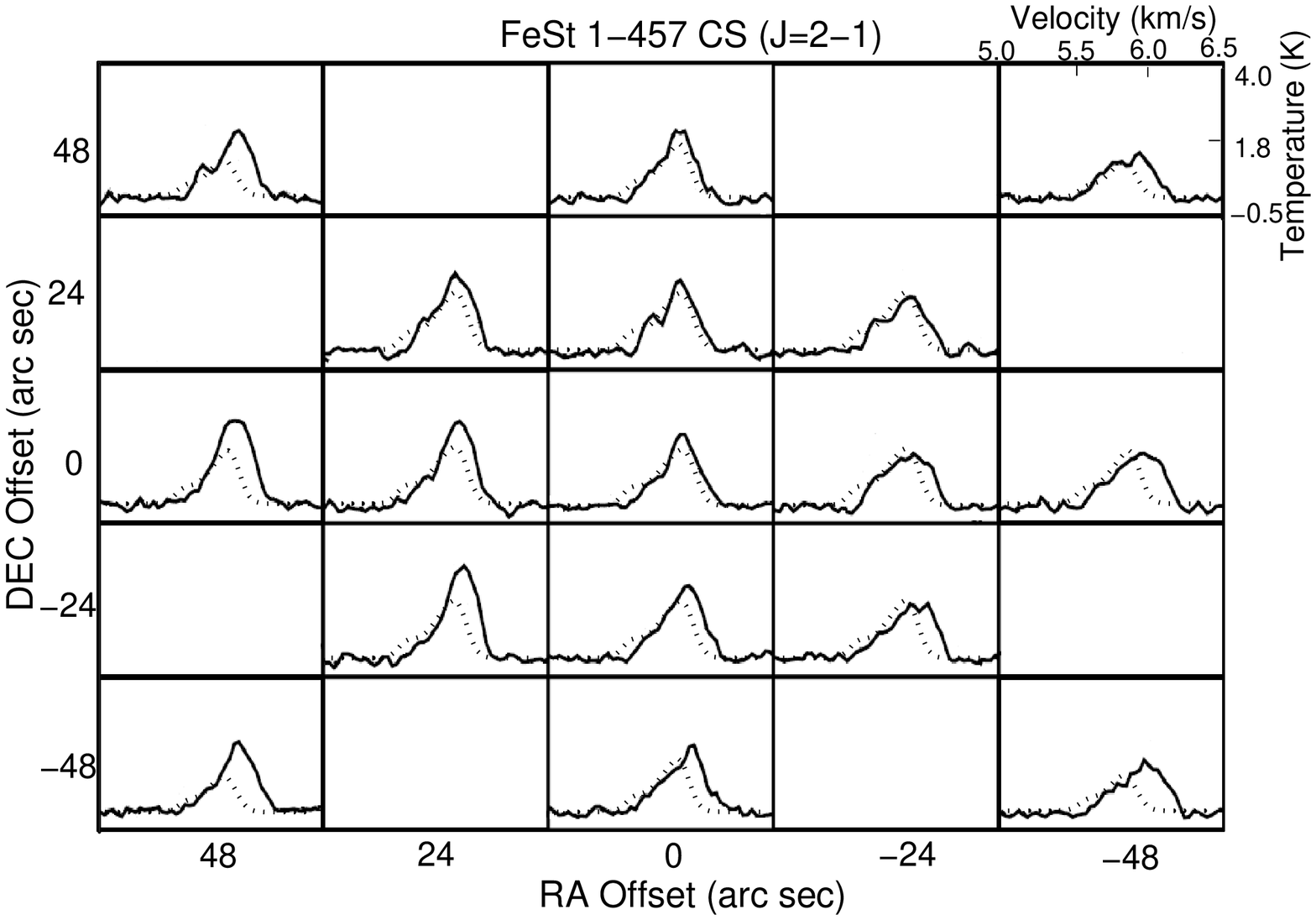,width=9cm,clip=}
\epsfig{figure=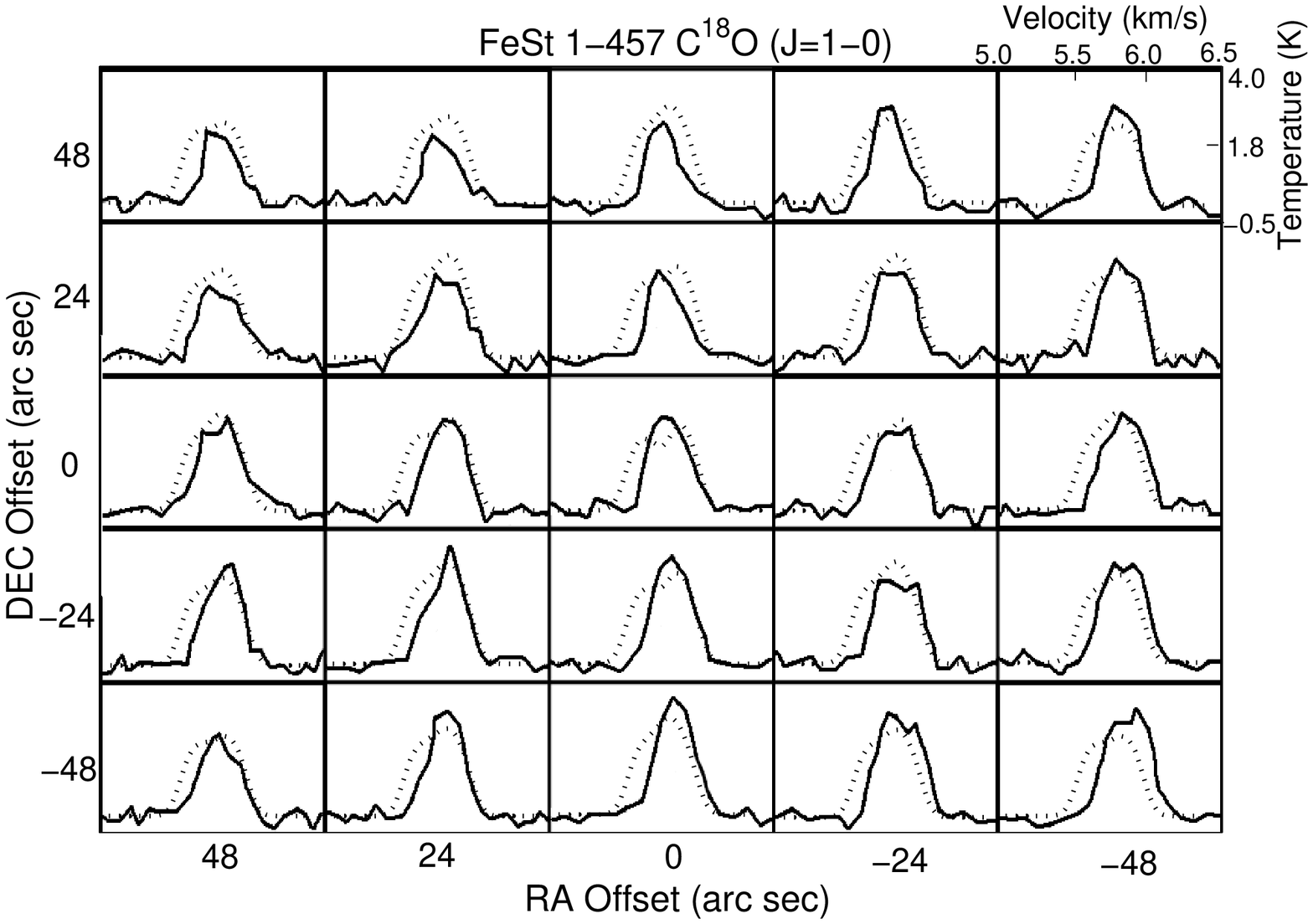,width=9cm,clip=}
\end{center}
\caption{Similar spatially-resolved grid-maps as shown in Fig.
  \ref{fig:hco}, but here for CS J$=2-1$ (top panel) and
  C$^{18}$O J$=1-0$ (bottom panel) molecular spectral line
  transitions.
The receding velocity of cloud core FeSt 1-457 along the
  line of sight is estimated as 5.75 km s$^{-1}$.
Spectral line fits are done using the same underlying polytropic
  EECC shock dynamic model described in the text, with step
  abundance profiles summarized in Table 1.
  \label{fig:csfit}
  }
\end{figure}

%
\begin{figure}
\begin{center}
\epsfig{figure=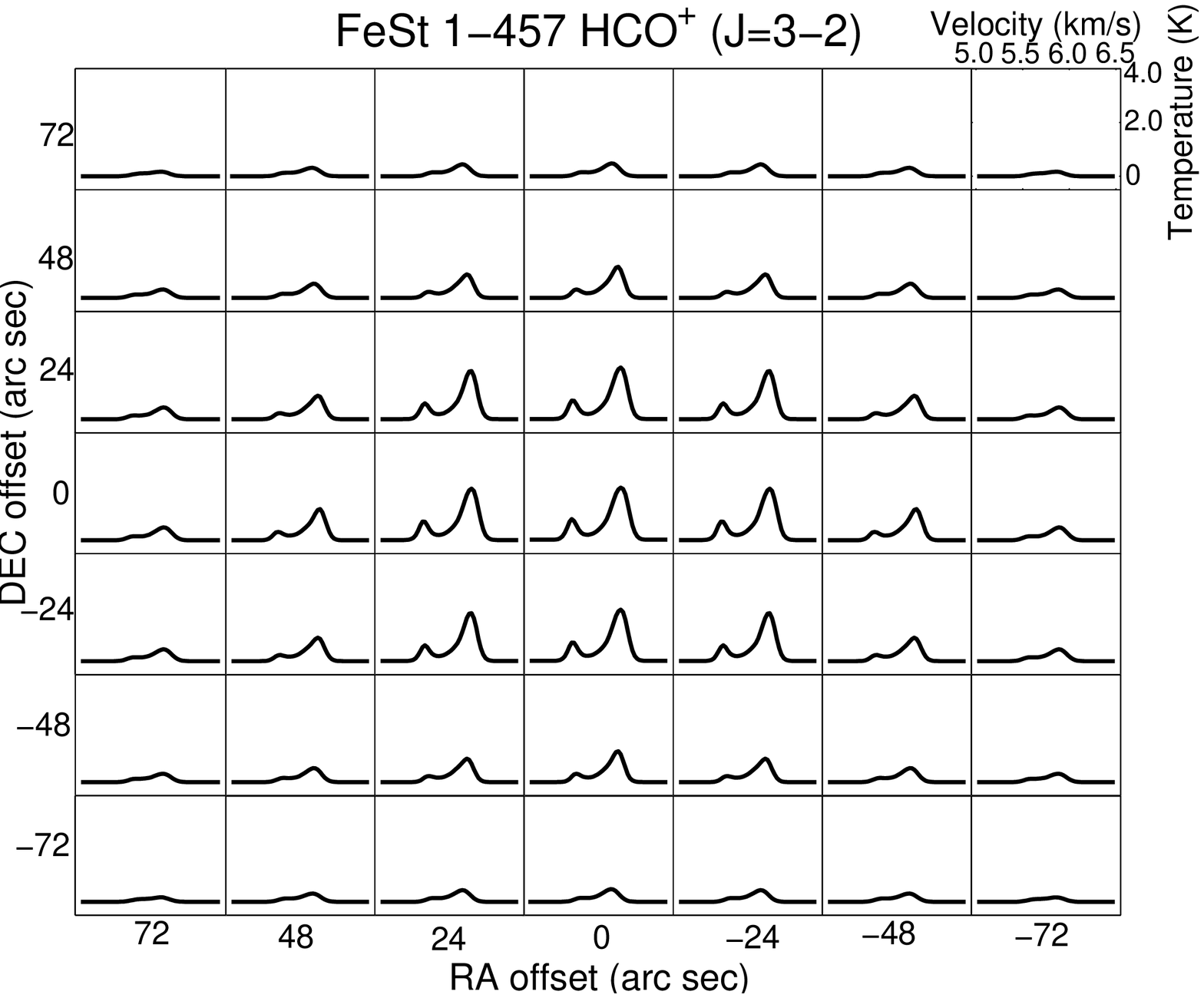,width=9.5cm,clip=}
\\
\epsfig{figure=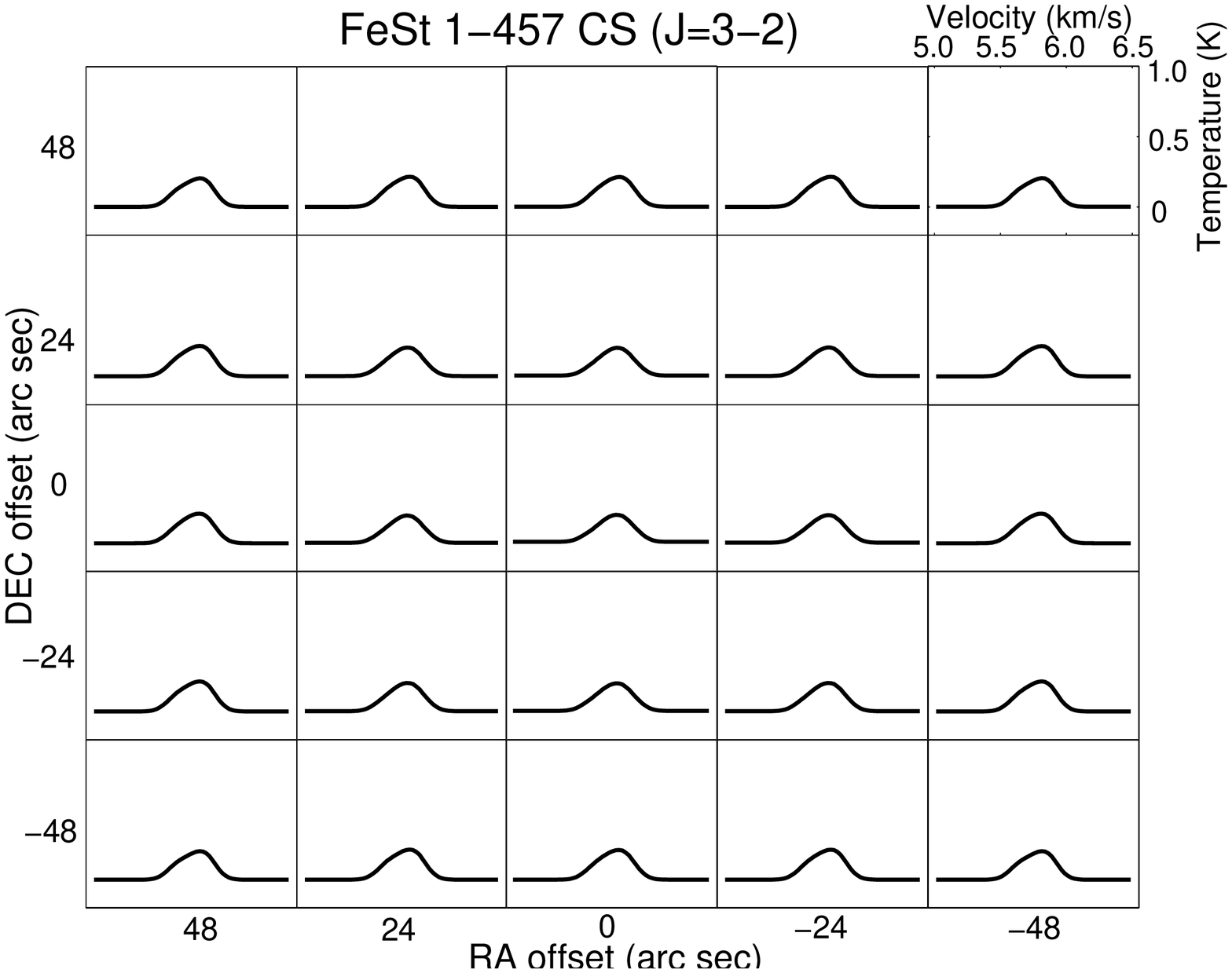,width=9.0cm,clip=}
\end{center}
\caption{Theoretical model predictions for the molecular
  transitions HCO$^+$ J$=3-2$ (top panel) and CS $J=3-2$
  (bottom panel) in spatially-resolved grid-maps
  of cloud core FeSt 1-457 on the basis of our best-fit EECC
  hydrodynamic shock model (see Section 3.2 for model details).
Each grid has the velocity as the abscissa (with the line-of-sight
  receding velocity 5.75 km s$^{-1}$ of FeSt 1-457) and the
  antenna temperature (for a 30m IRAM telescope beam efficiency
  of 0.75) as the ordinate.
Model spectral profiles are convolved with a typical beam width
  of $\sim 24''$.
The spatially-resolved grid-map is centered at the right ascension
  $\alpha_{J2000}=17^{\rm h} 35^{\rm m} 47.5^{\rm s}$ and the
  declination $\delta_{J2000}=-25^\circ 33' 2.0''$, with a grid
  spacing of $24''$.
The predication of HCO$^+$ J$=3-2$ emissions cover a maximum area of
  $168''\times 168''$; while for CS $J=3-2$ emissions, the spatial
  distribution is $120''\times 120''$.
Molecular abundance information are summarized in Table 1.
\label{fig:hco32}}
\end{figure}


Meanwhile, spatially-resolved CS (J$=2-1$) and C$^{18}$O (J$=1-0$)
  spectral line profiles are also computed using the RATRAN code
  given the same underlying EECC shock dynamic model and fitted with
  observational data of FeSt 1-457 shown in Fig. \ref{fig:csfit}.
The CS and C$^{18}$O abundance ratios are also prescribed
  as step functions as listed in Table 1.
In running the RATRAN code for radiative transfer calculations,
  molecular transition data of HCO$^+$, CS and C$^{18}$O are all
  taken from the Leiden Atomic and Molecular Database
  \citep[LAMDA;][]{schoier2005}.
The reasonable fits of CS J$=2-1$ and C$^{18}$O J$=1-0$ model
  molecular line profiles using the same EECC shock dynamic
  model to the observed corresponding line profiles (see Fig.
  \ref{fig:csfit}) confirm the viability of the underlying
  EECC shock model.
In comparison with spatially-resolved spectral line
  profiles of HCO$^+$ J$=1-0$, the observed CS J$=2-1$ and
  C$^{18}$O J$=1-0$ spectral line profiles show less asymmetric
  features because of the relatively lower optical depths in
  these two molecular line transitions.\footnote{As discussed in
  \citet{gaolou2010}, lower opacities would give rise to less
  asymmetries in molecular line profiles, thus not effective
  indicators of cloud hydrodynamics in star-forming clouds.}
Both Figs. \ref{fig:hco} and \ref{fig:csfit} show somewhat notable
  deviations between model spectral line profiles and those observed
  at grids in the lower right of the map, which is most likely
  associated with non-spherical symmetry of any realistic
  astrophysical systems.
We may infer from these complementary molecular line model
  fittings, though with less asymmetries in line profiles, that
  the spherically symmetric EECC model represents a reasonable
  description to this particular star-forming cloud.
It is expected that more structural complexities on
  smaller scales (e.g. Redman et al. 2004 for rotations
  and Carolan et al. 2008 for bipolar outflows
  etc.) can complicate such radiative transfer fittings.

As testable theoretical predictions for future telescope
  observations of the HCO$^+$ J$=3-2$ and CS J$=3-2$
  emission lines from the cloud core FeSt 1-457, we
  show our model calculation results as predictions
  in Fig. \ref{fig:hco32}.
Besides that all these predictions are based on the same EECC
  hydrodynamic shock model described in Section 3.2, abundance
  profiles of HCO$^+$ and CS remain also the same as what we
  have used to calculate HCO$^{+}$ J$=1-0$ and CS J$=2-1$
  molecular line transitions (see Table 1).
Molecular transitions HCO$^+$ J$=3-2$ and CS J$=3-2$ emissions
  have higher excitation temperatures than those for HCO$^{+}$
  J$=1-0$ and CS J$=2-1$ emissions, respectively
  \citep[e.g.][]{schoier2005}.
Therefore in the environment temperature of only $\sim 10$~K in
  a cloud, they are less excited and thus have lower emission
  intensities as shown in Fig. \ref{fig:hco32}.
The intensities of CS J$=3-2$ emissions are even weaker because
  of a larger central abundance hole presumed (see Table 1).
Because the abundance hole is large, less higher temperature
  cloud regions will contribute to the line emission, which
  makes the CS J$=3-2$ emission very weak.
This leads the conclusion that for certain molecules, the radius
  of the abundance hole will affect the intensity contrast between
  its transition lines with higher and lower excitation temperatures.
We may also learn from Fig. \ref{fig:hco} and Fig.
  \ref{fig:hco32} that optically thin molecular transition lines
  (here HCO$^+$ J$=3-2$) show less asymmetries in line profiles
  than optically thick transition lines (here HCO$^+$ J$=1-0$).
That is why optically thick lines are more sensitive diagnostics
  of hydrodynamic structures in molecular clouds.

\subsection{Column Density and Dust Continuum Profiles}

%
\begin{figure*}
\begin{center}
\epsfig{figure=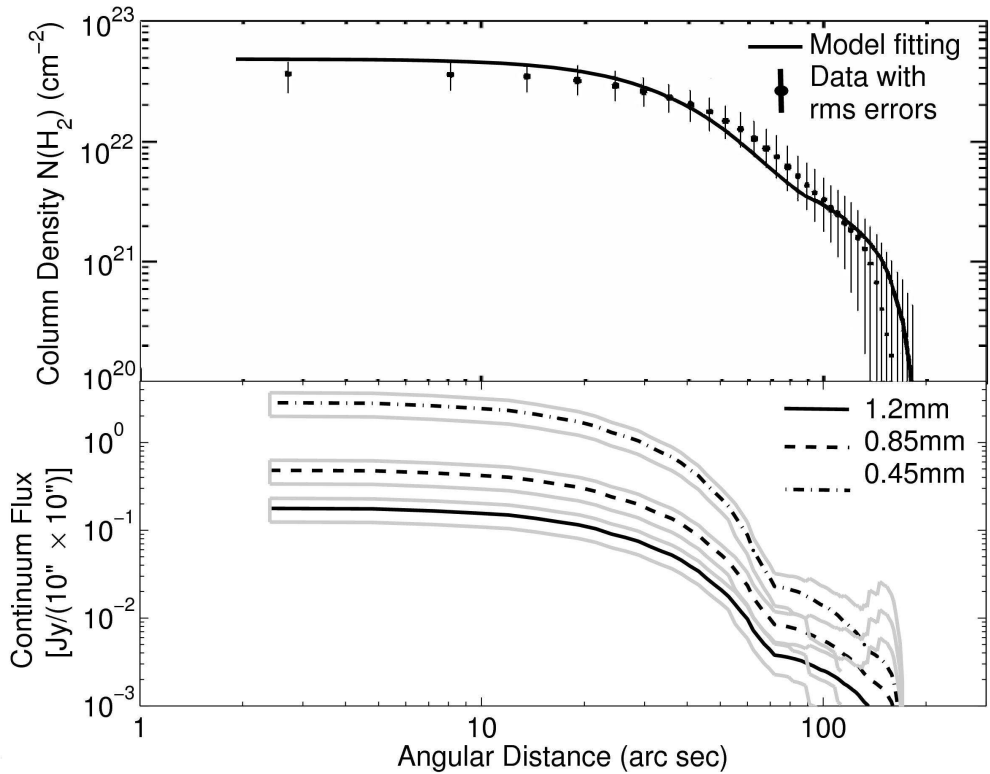,width=15cm,clip=}
\end{center}
\caption{Radial profiles of column number density (top panel)
  and predicted (sub)millimetre continuum emissions (bottom
  panel) are shown here.
The top panel compares the radial profile of column number
  density derived from the EECC hydrodynamic shock model (solid
  curve) and the data with error bars inferred from dust extinction
  observations \citep[][dots with errors]{kandori2005}.
Error bars are root-mean-square (RMS) scatters of data points
  within each annulus.
The radial profile derived from the EECC hydrodynamic shock
  model is convolved with the observational spatial resolution
  of $\sim 30''$.
The same EECC hydrodynamic shock model that fits molecular
  spectral line profiles in Fig. 1 also offers a fair fitting
  (within RMS errors) to the radial profile of column density.
The bottom panel shows continuum emission intensities at
  three wavelengths of 1.2~mm, 0.85~mm and 0.45~mm, which
  are predicted here for future observational tests.
These (sub)millimetre continuum profiles are derived using the
  same underlying EECC shock dynamic model, with dust opacities
  given by \citet{ossenkopf1994} in the case
  of bare ice mantles and $\sim 10^5$ yr growth, and convolved
  with a typical spatial resolution of $\sim 10''\times 10''$.
Three pairs of gray curves along the three predicted continuum
  profiles show relevant ranges of uncertainties in the
  (sub)millimetre continuum emissions, which are estimated
  according to RMS errors in the column number density
  observation, using the relationship $I_\nu\propto\rho^{1.2}$.
Abrupt turns around $\sim 70''$ signal shock discontinuities
  and might be detected with high enough spatial resolutions.
  \label{fig:continuum}
  }
\end{figure*}

The dust extinction observation of FeSt 1-457 of
  \citet{kandori2005} represents an important observational
  constraint to the model description of this molecular cloud core.
We infer from Fig. 2 of \citet{kandori2005} that the cloud core
  FeSt 1-457 appears grossly spherical, yet with slight asymmetry
  towards the lower right corner.
This asymmetry in the dust extinction observation coincides with
  the differences shown in spectral line profile model fittings
  as shown in Figs. \ref{fig:hco} and \ref{fig:csfit}.
All these available observations might imply that there exists
  some kind of beamed outflows towards the lower right direction
  in FeSt 1-457, in addition to an almost spherically symmetric
  cloud core.
As the deviation of molecular lines in this part of the cloud core
  is toward red-shifted in the order of $\sim 0.1\ {\rm km~s^{-1}}$
  (see Figs. \ref{fig:hco} and \ref{fig:csfit}), we estimate that
  a beamed slow outflow may recede from us with a typical speed of
  $\sim 0.1\ {\rm km~s^{-1}}$.

As a further consistent model check, we compute the radial profile
  of column density using the same EECC shock dynamic model and
  compare it with that inferred from the dust extinction data
  for FeSt 1-457 \citep[][]{kandori2005}.
As the column density data has a spatial resolution of $30''$,
  our computed model profile is derived from the same EECC
  shock dynamic model with a convolution in $30''$ and shows
  a sensibly good fit (the top panel of Fig. \ref{fig:continuum}).
As the radial profile of column density is sensitive to the outer
  radius $R_{\rm out}$ of the molecular cloud core and to the
  cloud number density $N$, this fit identifies a proper outer
  radius $R_{\rm out}$ and the molecular hydrogen H$_2$ number
  density scale (see eqn (\ref{equ:scale2})).
The outer radius of FeSt 1-457 used in the column density fit is
  $R_{\rm out}=1.8\times 10^4$~au, which is also used in molecular
  spectral line profile fits and appears consistent with that from
  the optical images \citep[][]{aguti2007} at the same time.
In short, the inferred variation trend of column density offers
  a constraint on a proper EECC shock dynamic phase with a grossly
  self-consistent radial density profile.

In previous publications
  \citep[e.g.][]{alves2001,harvey2003,kandori2005}
  regarding the column density structure of molecular clouds,
  researchers normally use a power-law radial density distribution
  or a static Bonnor-Ebert sphere \citep{bonnor1956,ebert1955} etc.
  to describe density distributions within observed molecular clouds.
These empirical/theoretical models also appear to provide
  reasonable fits to the observationally inferred radial column
  density distributions, but none of them have been shown to
  successfully account for the observed molecular spectral line
  profiles from those molecular clouds, except that additional
  parameterized conditions are introduced \citep[e.g.][]{tafalla2006}.
In contrast, our theoretical cloud models involving proper
  hydrodynamic flows and with self-consistent radial temperature
  variations can give rise to asymmetric molecular line profiles
  \citep{gao2009,gaolou2010}.
According to the Bonnor-Ebert sphere model fits to a group of
  cloud cores in \citet{kandori2005}, most of those best-fit
  static spheres are actually unstable Bonnor-Ebert spheres,
  implying that these molecular cloud cores are most likely
  undergoing hydrodynamic evolution in reality.
Another data analysis by \citet{li2007}
  on massive quiescent cores in the Orion molecular cloud,
  gives a similar result that most of the cloud cores are
  unstable, if modelled as static Bonnor-Ebert spheres.
All these results hint that we do need a theoretical model
  formulation for the hydrodynamic evolution of cloud cores.
That is, the spectral line profiles and dust emission data
  fitting are not inconsistent with our EECC dynamic shock model.
Essentially similar problems also exist in the model fits of dust
  continuum emission observations described immediately below.
Therefore to be consistently constrained by all available
  observations, we explore the polytropic EECC hydrodynamic
  shock model here for the molecular cloud core FeSt 1-457.

Radial profiles of (sub)millimetre continua at three wavelengths
  can be also computed using the RATRAN code to further check the
  consistency for the density and temperature radial structures
  of cloud cores.
For these testable observations
  \citep[e.g.][for other cloud systems]{shirley2000,shirley2002},
  we specifically predict such continuum radial profiles at
  three wavelengths of 1.2~mm, 0.85~mm and 0.45~mm using the
  same underlying EECC hydrodynamic shock model for FeSt 1-457.
Modelling such dust continuum emissions as optically thin with
  a constant dust opacity $\kappa_{\nu}$ in $r$ at a specified
  frequency $\nu$, the observed intensity at an impact parameter
  $b$ is \citep[][]{adams1991,shirley2002} given by
  \begin{equation}
  I_{\nu}(b)=2\int_b^{R_{\rm out}}B_{\nu}(T_{\rm
  d})\frac{\kappa_{\nu}\rho(r)r}{(r^2-b^2)^{1/2}}dr\ ,
  \label{equ:continuum}
  \end{equation}
where $T_{\rm d}(r)$ is the dust temperature taken to be the
  same as the H$_2$ gas temperature $T(r)$, $\rho(r)$ is the
  mass density of H$_2$ molecules and $B_{\nu}(T_{\rm d})$
  is the Planck function.
The opacity $\kappa_{\nu}$ is that of \citet{ossenkopf1994}
  dust model for bare ice mantles with a growth time of
  $\sim 10^5$ yr.
As usual, a H$_2$ gas to dust mass ratio 100 to 1 is adopted
  in our model computations.
With the dust temperature and density profiles from the EECC
  hydrodynamic shock model, we obtain radial profiles for
  1.2~mm, 0.85~mm and 0.45mm continuum intensities.
Results shown in the bottom panel of Fig. \ref{fig:continuum} are
  convolved with a telescope beam size of $\sim 10''\times 10''$
  and a beam efficiency of $\sim 1$.
Error ranges of millimetre continuum intensities are estimated by
  the root-mean-square (RMS) errors in the column density observation
  (the upper panel of Fig. \ref{fig:continuum}), through the relation
  $I_\nu\propto\rho^{1.2}$.
The abrupt `kink' at $\sim 70''$ from the centre reveals shock
  discontinuities of density and temperature in the EECC shock model
  (see the middle and bottom panels in Fig. \ref{fig:dynamics}),
  and should be detectable with high enough spatial resolutions.

\section{Conclusion and Speculations}

We have presented model results pertinent to three complementary
  observational aspects towards the dark molecular globule FeSt
  1-457; fairly reasonable fittings to asymmetric `red profiles'
  and to the column density radial profile derived from dust
  extinction data indicate that this dark globule most likely
  involve an early EECC shock phase of dynamic cloud evolution,
  while our predictions for spatially-resolved grid-maps for
  HCO$^{+}$ J$=3-2$ and CS J$=3-2$ respectively as well as the
  three predicted millimetre continuum radial profiles can be
  specifically tested by future observations.
The choice of our model parameter set is determined by an
  extensive exploration and known quantitative as well as
  qualitative constraints.
Slight adjustment around this chosen model parameter set is
  allowed for a refined improvement of fitting.
Although without a rigorous proof, it would be surprising that
  a dramatically different model parameter set could reproduce
  all the available observational data so far.
In addition to the spherical global EECC dynamics, line
  profile and extinction observations might also suggest a
  slow beamed outflow receding from us at a typical speed of
  $\sim 0.1 {\ \rm km~s^{-1}}$ towards the lower right direction
  in the grid-maps.
The inferred underlying EECC hydrodynamic shock model
  provides physical parameters for FeSt 1-457 cloud core.
The central protostellar mass blob and the total mass of the cloud
  core are $M_0\sim 0.020M_\odot$ and $M_{\rm tot}\sim 2.93M_\odot$,
  respectively; the central protostellar mass accretion rate
  is $\dot{M}_0\sim 0.69\times 10^{-7}M_\odot~{\rm yr^{-1}}$
  (see eq (\ref{equ:massaccretion}));
  the dynamical timescale of the cloud core FeSt 1-457 is
  estimated as $t\sim 2.8\times 10^5~{\rm yr}$ accordingly
  (see expression (\ref{equ:scale2})).
As an expanding shock travels through the molecular cloud
  core, abundance differences of certain tracer molecules
  (e.g. CH$_3$OH and SiO) across the shock front
  \citep[e.g.][]{jorgensen2004} could be detectable.

As a strong supporting evidence to the present case of FeSt 1-457,
  a parallel theoretical model analysis on another star-forming
  molecular cloud core L1517B has also been conducted and
  completed recently \citep{FuGaoLou2010}.
Several molecular line transitions of L1517B also manifest red
  asymmetric line profiles.
On the basis of a shocked EECC similarity hydrodynamic model,
  several single point central molecular emission lines of
  HCO$^+$($1-0$), HCO$^+$($3-2$), H$^{13}$CO$^+$($1-0$),
  DCO$^+$($3-2$), H$_2$CO($2_{12}-1_{11}$),
  H$_2$CO($2_{11}-1_{10}$), CS($2-1$), CS($3-2$), SO($23-12$),
  and SO($34-23$) from the source L1517B [observed by 13.7~m
  telescope of the Five College Radio Astronomical Observatory
  (FCRAO) and IRAM 30~m telescope] are simultaneously fitted
  in an overall consistent manner.
Taking into account of the fact that spatially-resolved HCO$^+$
  (J$=1-0$) emission line profiles [observed by Delingha 13.7m
  telescope of the Purple Mountain Observatory (PMO)], as well
  as 1.2~mm and 850~$\mu$m continuum emissions (observed by
  IRAM 30~m and JCMT respectively) can be consistently fitted
  by the same underlying EECC hydrodynamic shock model, we are
  fairly confident that the cloud core L1517B appears to be
  another cloud core that evolves in the EECC shock phase.
Similar to what we have done for FeSt 1-457 here, predictions of
  several other spatially-resolved molecular transition lines,
  the radial profile of 450~$\mu$m continuum emission, and
  visual extinction (all based on the same EECC hydrodynamic
  shock model) are also presented for further observational
  tests of the model description of L1517B (see
  \citet[][]{FuGaoLou2010} for details).

We speculate that other starless cores of grossly spherical
  molecular clouds with `red profiles' might also involve
  EECC hydrodynamic shock phase in general.
More sources need to be search and established by systematic
  model analysis and data comparisons.
The existence of global EECC hydrodynamic shock at early phases
  would affect the formation of protostars and their surroundings,
  including estimates for star formation rates and initial mass
  functions that need to be further explored.

\section*{Acknowledgments}

This research was supported in part by Tsinghua Centre for
  Astrophysics (THCA), by the National Natural Science
  Foundation of China (NSFC) grants 10373009, 10533020
  and 11073014 at Tsinghua University, and by the Yangtze
  Endowment and the SRFDP 20050003088 and 200800030071
  from the Ministry of Education at Tsinghua University.

\end{document}